\documentclass{elsart}
\usepackage{epsfig}
\usepackage{amsmath}
\usepackage{amsbsy}

\newcommand{\be}{\begin{equation}}
\newcommand{\ee}{\end{equation}}
\newcommand{\bea}{\begin{eqnarray}}
\newcommand{\eea}{\end{eqnarray}}

\newcommand{\LB}[1]{\label{#1}}
\newcommand{\R}{\mbox {$I\kern-4pt R $}}
\begin{document}
\bibliographystyle{unsrt}\baselineskip 15pt
\begin{frontmatter}

\title{Instabilities and Spatio-temporal Chaos of Long-wave Hexagon
Patterns in Rotating Marangoni Convection} 
\author{A. M. Mancho}\footnote{Previously at: Departamento de F\'{\i}sica y
Matem\'atica Aplicada,   Facultad de Ciencias, Universidad de Navarra,
31080 Pamplona, Navarra, Spain. \\Current address: School of Mathematics,
University of Bristol, University Walk, Bristol BS8 1TW, United Kingdom}
 \address{Centro de Astrobiolog\'{\i}a 
(Associated Member of NASA Astrobiology Institute)
       CSIC-INTA, Ctra. Ajalvir km. 4, 28850 Torrej\'on de Ardoz, Madrid, Spain}
\author{ Hermann Riecke and Fil Sain}\footnote{Current address: 
The CNA Corporation, 4401 Ford. Ave, Alexandria, VA 22302}
\address{ Department of Engineering Science and Applied Mathematics,
        Northwestern University, 2145 Sheridan Road, Evanston, IL 60208, USA }
\maketitle

\begin{abstract}
We consider surface-tension driven convection in a rotating fluid layer. For
nearly insulating boundary conditions we derive a long-wave equation for
the convection planform. Using a Galerkin method and direct numerical
simulations we study the stability of the steady  hexagonal patterns with
respect to general side-band instabilities. In the presence of rotation steady
and oscillatory instabilities are identified. One of them leads to stable,
homogeneously oscillating hexagons. For sufficiently large rotation rates the 
stability balloon closes, rendering all steady hexagons unstable and leading to
spatio-temporal chaos.
\end{abstract}

\end{frontmatter}
\newpage

{\it In the study of the formation of patterns the origin of spatial and temporal
disorder  is an important question. In recent years quite some progress has
been made in  the understanding of various states of spatio-temporal chaos
that arise from  stripe-like patterns, as they are found in the  convection of
fluids heated from below,  or from spatially homogeneous oscillations. Much
less is known about spatio-temporally chaotic  states that are based on
hexagonal patterns. We therefore investigate the classic
surface-tension-driven Marangoni convection in a modification in which the
system  is rotated around a vertical axis. In buoyancy-driven roll convection
such a rotation is known to  induce chaotic dynamics through the
K\"uppers-Lortz instability. In order to obtain a systematic reduced
description of the fluid system we consider the case of long-wave
convection and derive the appropriate order-parameter equation. Its stability 
analysis reveals regimes in which hexagons of all wavenumbers are unstable
leading to a spatio-temporally chaotic state. In addition, homogeneously
oscillating hexagon states are found. }

\section{Introduction}
\LB{sec:intro}
 
In the classic, buoyancy-driven Rayleigh-B\'enard convection
rotation of the system has been found to lead to very interesting new
dynamic patterns. The origin of the dynamics is the K\"uppers-Lortz
instability \cite{KuLo69}, which renders all straight roll patterns
unstable with respect to a set of rolls rotated with respect to the
original ones. It  occurs above a critical rotation rate. In simple
low-dimensional models the instability is  found to a heteroclinic
cycle connecting roll patterns rotated with respect to each other by
$120^o$ \cite{BuCl79,BuHe80}. In experiments (e.g.
\cite{ZhEc91,HuPe98}) as well as in numerical computations of the
Navier-Stokes equations (e.g. \cite{HuPe98}) and of model equations
(e.g. \cite{CrMe94}) more complex behavior is found. Patches of
convection rolls of various orientations arise and the roll orientation 
changes through the propagation of the fronts separating them or by
the nucleation of rolls with a new orientation. 

The other classic convection system is surface-tension-driven
Marangoni convection of a liquid layer in contact with another liquid
or a gas layer (for a recent review see \cite{ScNe01}). Two different
instability mechanisms have been identified that drive this type of
convection: a short-wave instability, which occurs independent  of
any deflections of the interface \cite{Pe58} and a long-wave
instability, for which  deformations of the interface are essential
\cite{ScSt64,Sm66}.  Not too far from threshold, the short-wave
instability typically leads to convection patterns with hexagonal
planform (e.g. \cite{JuBu00,EnSw00}), while the long-wave instability induces 
isolated dry spots or high spots \cite{VaSc95,VaSc97}. 
The stability of the hexagonal short-wave pattern with respect to 
sideband instabilities (the 'stability balloon') has been determined
in weakly nonlinear \cite{EcPe01} and in fully nonlinear numerical approaches 
\cite{Be93}. Further above onset the short-wave hexagon  pattern is 
often observed to undergo a  transition to squares
\cite{NiTh95,EcBe98,ScVa99,JuBu00,ToMo00,EnSw00}. If the
mean surface tension of the fluid is not too large the short-wave
convection  mode has to be coupled to a weakly damped long-wave
mode that corresponds to the local layer height of the fluid
\cite{GoNe94,GoNe95}. A different type of long-wave instability has
been identified theoretically for the case when the fluid layer is
bounded above and below by poor thermal conductors \cite{Pe58}.
The mechanism for this instability is the same  as for the short-wave
instability. Due to the poor conductivity the critical wavenumber  is,
however, shifted to small values.  Again one finds that near onset the
planform is usually hexagonal. It turns out that experimentally this limit
of long wavelengths is difficult to achieve. Theoretically, however, it
is a very  useful limit since it allows to derive simplified evolution
equations systematically  from the basic fluid equations without the
restriction to small amplitudes, although restrictions on the control
parameter remain \cite{Si82,Kn90,GoNe95b}. Moreover, compared to the 
weakly nonlinear Ginzburg-Landau equations the long-wave
equations have the additional great advantage of preserving the
isotropy of the system. This will play an important role in the present
paper.

With the interesting behavior ensuing from the K\"uppers-Lortz instability, it is 
natural to ask whether rotation has also an interesting impact on hexagonal
convection patterns. Within the general context of small-amplitude 
expansions and Ginzburg-Landau equations this question has been
addressed previously
\cite{Sw84,So85,GoKn92,Ri94,MiPe92,EcRi00,EcRi00a,EcRi00b}.  These analyses
were aimed at  a general investigation of hexagonal patterns in rotating
systems and apply  also to weakly non-Boussinesq Rayleigh-B\'enard
convection. It was found that in a rotating system a new state of oscillating
hexagons can appear. This transition  replaces then the  transition from
hexagons to rolls that is obtained in non-Boussinesq Rayleigh-B\'enard 
convection. In Marangoni convection this transition does not occur and no
oscillating hexagons of this type are expected. Instead, in the absence of
rotation a transition to squares is found \cite{NiTh95,EcBe98,ScVa99}. The
effect of rotation on  this transition has not been studied yet. For the  usual,
steady hexagons the weakly nonlinear analysis indicates the possibility of
persistent chaotic dynamics coexisting with the steady hexagons for the
same parameter values \cite{EcRi00}.  However, these dynamics cannot be
captured by the Ginzburg-Landau equations employed. To make reliable
statements about this chaotic state, which is characterized by a  spatially
isotropic Fourier spectrum, equations have to be studied that preserve the
isotropy of the system. A first step in this direction was  the study of a
Swift-Hohenberg model, which was generalized to capture the breaking of
the chiral symmetry that characterizes the rotation. A detailed analysis  of the
resulting equation, which was not derived from any physical  system,
revealed that steady hexagons of all wavenumbers can become unstable
already for amplitudes below the transition to the oscillating hexagons,
leaving a range  in the control parameter over which no simple periodic
solution is stable \cite{SaRi00} and in which chaotic dynamics are found in
numerical simulations.

Motivated by the previous results of weakly nonlinear approaches to 
hexagons in rotating systems, our goal for the present paper is to obtain a
systematic description of Marangoni  convection in a rotating system that is
not limited to small amplitudes and that preserves the isotropy of the system.
We therefore turn to the long-wave limit that is obtained for poorly
conducting top and bottom boundaries and extend the derivation of
\cite{Si82,Kn90} to the  case of a rotating system. The poor heat conductivity
is expressed in terms of a small Biot number $Bi$. While the true limit of long
wavelengths ($Bi\ll1$) may not be attainable in experiments \cite{VaSc97}
we expect to gain valuable insight into the system  for moderate wavelengths
by expanding in $Bi$. The form of the resulting long-wave equation is the same
as that found for buoyancy-driven convection with poorly conducting boundaries \cite{Co98}
and for buoyancy-driven convection in binary mixtures with positive separation ratio
\cite{Kn89}. In the latter case the critical wavenumber goes to 0 at a critical
value of the separation ratio. In both cases quadratic terms arise only if the
vertical symmetry is broken by imposing different boundary conditions at the
top and the bottom plate (free-slip vs. no-slip or fixed temperature vs. fixed 
heat flux). The focus of \cite{Co98} is the effect of rotation
on the system. In contrast to our work, however, there the competition
between strictly periodic roll and square planforms is considered. The
work is performed analytically in the weakly nonlinear regime, supplemented
by a few simulations of a small system.

In sec.\ref{sec:problem} we introduce the  basic fluid
equations and associated boundary conditions. The general linear stability 
analysis that identifies the long-wave regime is performed in
sec.\ref{sec:linstab}. The long-wave equation is derived in
sec.\ref{sec:longwave}. The stability and dynamics of the hexagon patterns is
then studied numerically in sec.\ref{sec:num}. The main results of the simulations
are periodic and chaotic dynamics arising from secondary instabilities of the
hexagons as well as from a resonant mode interaction.

\section{Formulation of the problem}
\LB{sec:problem}
\subsection {Basic equations}
We consider a thin viscous and incompressible fluid layer heated from below
with a free surface and  rotating around a vertical axis. The mean surface
tension is assumed to be large enough to suppress surface deformations
even in the presence of rotation. The Navier-Stokes and the heat equation
then read
  \bea
  {\rho}\left( \partial_t  + {\bf v} \cdot \nabla \right) {\bf v}
 &=& - \nabla p
+ \frac{\rho}{2} \nabla | \Omega {\bf e}_z \times {\bf r} |^2 + 2  \rho
\Omega {\bf e}_z \times {\bf v} +
\eta \Delta {\bf v}-\rho g {\bf e}_z, \LB{e:fev}\\
  \nabla \cdot {\bf v} &=& 0, \LB{e:fediv}\\
 \left( \partial_t  + {\bf v} \cdot \nabla \right) T &=&  
\kappa \nabla^2 T. \LB{e:feT}
\eea
In this paper we focus on fluid layers that are sufficiently thin that buoyancy
can  be ignored. The density is then taken to be constant, $\rho=\rho_0$. 
The dynamic viscosity $\eta$ is related to the kinematic viscosity $\nu$
through $via$ $\nu=\eta/\rho_0$  and $\kappa$ is the thermal diffusivity. The
equations in  dimensionless form can be obtained if the space,  time, angular
velocity, temperature, pressure  and velocity fields are respectively divided
by the constants $d$, $d^2/\kappa$, $\kappa/d^2$, $\Delta T$, $ \rho_0
\kappa \nu/d^2$ and $\kappa/d$. Here $d$ is the thickness of the fluid layer
and $\Delta T$ is the temperature difference across the liquid layer in the
conductive state. They are written as follows,
\bea
\frac{1}{\mathcal{P}}\left( \partial_t  + {\bf u} \cdot \nabla \right) {\bf u}
& =& - \nabla p
+ \frac{\mathcal{P}}{2} \nabla |{\bf \Omega} \times {\bf r} |^2 + 2 
{\bf \Omega} \times {\bf u} +
\nabla^2 {\bf u}-{\mathcal {G}} {\bf e}_z,\LB{e:feva}\\
\nabla \cdot {\bf u} &=& 0,\LB{e:fediva}\\
\left( \partial_t  + {\bf u} \cdot \nabla \right) \theta &=&  
\nabla^2 \theta,\LB{e:feTa}
\eea
where ${\mathcal{P}}=\nu/\kappa$ is the Prandtl number and
${\mathcal{ G}}=d^3\rho g/( \eta \nu)$ is the Galileo number. In this
paper we consider the limit of infinite Prandtl number.

We eliminate the pressure from the Navier-Stokes equations (\ref{e:feva}) by
using the incompressibility condition (\ref{e:fediv}). Taking the curl and the
double curl of (\ref{e:fev}) we obtain \cite{MaNe93} 
  \bea 
\nabla^4 w - 2 \Omega \, \partial_z \zeta &=&0
  \LB{e:1},\\ 
\nabla^2 \zeta + 2 \Omega \,\partial_z w &=&0,
  \LB{e:2} 
\eea
where $w$ is the $z$-component of the velocity ${\bf u}=(u,v,w)$  and
$\zeta$ is the $z$-component  of the vorticity $\nabla \times {\bf u}$. We
also introduce the horizontal Laplacian 
$\Delta_H=(\partial^2_{xx}+\partial^2_{yy})$. Note that the rotation couples
the vertical velocity  and the vertical vorticity. Thus, while in the absence of
rotation the vertical vorticity vanishes in many solutions of interest, with
rotation the vertical vorticity is in general non-zero. 
 
 The equations for the $x$ and $y$-components of the velocity field are
obtained   considering that the velocity field ${\bf u}=(u,v,w)$ is expressed in  
terms of the poloidal and the toroidal stream functions $\phi$ and $\xi$ as
follows:   $ {\bf u}= (\partial_x \partial_z \phi + \partial_y \xi, 
 \partial_y \partial_z \phi - \partial_x \xi, -\Delta_H \phi)$.
Then the equations for  $u$ and $v$  are:
   \bea
   - \Delta_H u &=& \partial_x \partial_z w + \partial_y \zeta,\LB{e:3} \\
   - \Delta_H v &=& \partial_y \partial_z w - \partial_x \zeta.\LB{e:4}
\eea
 These equations together with (\ref{e:1})-(\ref{e:2}) are
 equivalent to (\ref{e:feva})-(\ref{e:fediva}) \cite{Kn90}.

\subsection{Boundary conditions}
In order to obtain the velocity boundary condition at the top free surface we
take into account thermocapillary effects and assume a linear dependence 
of the surface tension on the temperature,   $\sigma=\sigma_0-\gamma
(T-T_0)$. Furthermore, for sufficiently large mean surface tension and not
too high rotation rates we can take the  free surface to be flat.  The boundary
conditions at the upper free surface are then given by 
\bea
\rho \nu \left( \frac{\partial w}{\partial x} +  \frac{\partial u}{\partial z} \right)
&=& - \gamma \left(\frac{\partial T}{\partial x} \right)\LB {e:m1}, \\
\rho \nu \left(\frac{\partial w}{\partial y} + 
\frac{\partial v}{\partial z}\right)&=& -\gamma \left(\frac{\partial T}{\partial y} \right).
\LB{e:m2}
\eea
After taking derivatives of (\ref{e:m1}) and  (\ref{e:m2}) with respect to
$x$ and $y$, respectively, and addition of both expressions, one is led
to
\bea
\rho \nu \left(\frac{\partial^2 w}{\partial z^2}\right)& =& - \gamma \Delta_H  T.
\eea
Introducing the Marangoni number  $M=\gamma d \Delta T/(\rho \nu \kappa) $
the boundary condition can be written as
\bea
\frac{\partial^2 w}{\partial z^2} &= &M \Delta_H \theta \qquad \mbox{ at } z=1. \LB{e:5}
\eea

From the Marangoni equations (\ref{e:m1}-\ref{e:m2}) and from the underformability of the 
surface the other boundary conditions at the top are
\bea
w= \partial_z \zeta=0, \LB{e:9}&& {\rm at} \, z=1.
\eea
At the bottom  we take no-slip boundary conditions for the velocity
\cite{ViAc66,KaLe94,JoKu97,Co98},
\bea
 w=\zeta=\partial_z w=0,\LB{e:8} && {\rm at}\, z=0. \LB{e:bcbot}
\eea

The critical wavenumber at which convection sets in depends sensitively on the temperature
boundary conditions. Long wavelengths are obtained if the boundaries are poor conductors.
The Newton law at the bottom plate can be expressed as
\bea
{\rm k} \partial_z T_{fb} &=& h_1 (T_{fb}-T_b) \qquad \mbox{ at }z=0, \LB{e:nl}
\eea
where $T_{fb}$ is the temperature of the fluid at the bottom plate and
$T_b$ is the temperature of the plate at some distance below this
interface. The thermal conductivity of the liquid is given by k and
$h_1$ is the heat exchange coefficient at the bottom. We
non-dimensionalize the temperature field by setting $\theta=(T-T_{fbc})/\Delta T$, 
where  $T_{fbc}$ is the temperature of the fluid
in the conductive state at the bottom plate  and $\Delta T$ is again the temperature difference across
the layer  in the 
conductive state, i.e. $\Delta T = T_{fbc}-T_{ftc}$ with $ T_{ftc}$ the
upper temperature of the fluid  in  the conductive state. In terms of $\theta$
the boundary condition (\ref{e:nl}) is then given by 
\bea
\partial_z \theta &=& Bi_1 \theta -1 \qquad \mbox{ at } z=0 \LB{e:6}
\eea
with the Biot number $Bi_1 = h_1 d/{\rm k}$. Here we have made use of the conservation 
of heat flux at the interface, $h_1 (T_{fbc} - T_b)= - {\rm k}/d (T_{fbc} -T_{ftc})$.
In a similar way the temperature boundary condition at the top  is found to be
\bea
\partial_z \theta &=& - Bi_2 \theta -1 - Bi_2 \qquad \mbox{ at } z=1. \LB{e:7}
\eea
These expressions agree with those used in \cite{GoNe95a} when the limit of a single layer is taken 
(in their notation this limit corresponds to $\Theta_1=\Theta_2$, $\kappa = 1$ and $a=0$).

\section{Linear stability analysis}
\LB{sec:linstab}

In order to motivate the long-wavelength analysis we first discuss the
linear stability of the conductive solution. The normal modes of the
linearized system (\ref{e:feTa}-\ref{e:2}) together the boundary
conditions (\ref{e:5}-\ref{e:bcbot}), (\ref{e:6}-\ref{e:7}), take the form
\[
({\bf u}, \theta)=  (\hat{{\bf
u}}(z),\hat{\theta}(z)) e^{i{\bf q}\cdot{\bf r } + \Lambda t},
\]  
where $\Lambda$ is the  growth rate, ${\bf q}$ is the  wavevector and {\bf r}
is the position vector $(x,y)$.   Since we are interested only in
thermocapillary effects  the Marangoni number is the sole control  parameter.
At the critical value $M_c$ of the Marangoni number, determined by a
vanishing  real part of $\Lambda$, the conductive state first becomes
unstable.  If the growth rate is $\Lambda=0$ at $M_c$ the bifurcation is
steady, while if  $\Lambda = i \omega$ for some nonzero real frequency
$\omega$, the bifurcation is oscillatory.  In this work we examine only the
steady bifurcation. It is known that for sufficiently large rotation  rates an
oscillatory bifurcation may occur \cite{McFi69}. In the absence of rotation,
the linearized system (\ref{e:feTa}-\ref{e:2})  yields the critical Marangoni
number $M_c$     as a function of the wave number $|{\bf q}|$ and the Biot
numbers $Bi_{1,2}$
  \begin{eqnarray}
\lefteqn{{M} = 8{q} \left( {\vrule 
height0.44em width0em depth0.44em} \right. \! \!  \left( \! \, - 
 {q}^{2}\,{\rm cosh}q +  {q}\,{\rm cosh}q^{2}
\,{\rm sinh}q\, \!  \right) \, (Bi_1+ Bi_2)}         \nonumber \\
 & & \mbox{} +  \left( \! \, - {q}\,{\rm sinh}q + {\rm 
cosh}q^{3} - {\rm cosh}q\, \!  \right) \,  
Bi_1 Bi_2  \nonumber \\
&& + {q}^{2}\,{\rm cosh}q^{3}  
     - {q}^{2}\,{\rm cosh}q - {q}^{3}\,{\rm sinh}
q \! \! \left. {\vrule height0.44em width0em depth0.44em}
 \right) / A,   
\LB{e:linstab} 
\end{eqnarray}
where,
 \begin{eqnarray}
 & & {\rm A} =  
   \left( \! \,{\rm cosh}q^{2}\,{\rm sinh}q - 
{\rm sinh}q - {q}^{3}\,{\rm cosh}q\, \!  \right) 
\,  Bi_1  + {q}^{3}\,{\rm cosh}q \\
 & & \mbox{} - 2\,{q}^{2}\,{\rm sinh}q + {q}\,{\rm cosh}q^{3} 
- {q}\,{\rm cosh}q - {q}^{4}\,{\rm sinh}q \! \! \left. {\vrule height0.44em width0em depth0.44em}
 \right). 
\LB{e:linstabA}
\end{eqnarray}
\begin{figure}[tbp]
  \epsfig{file=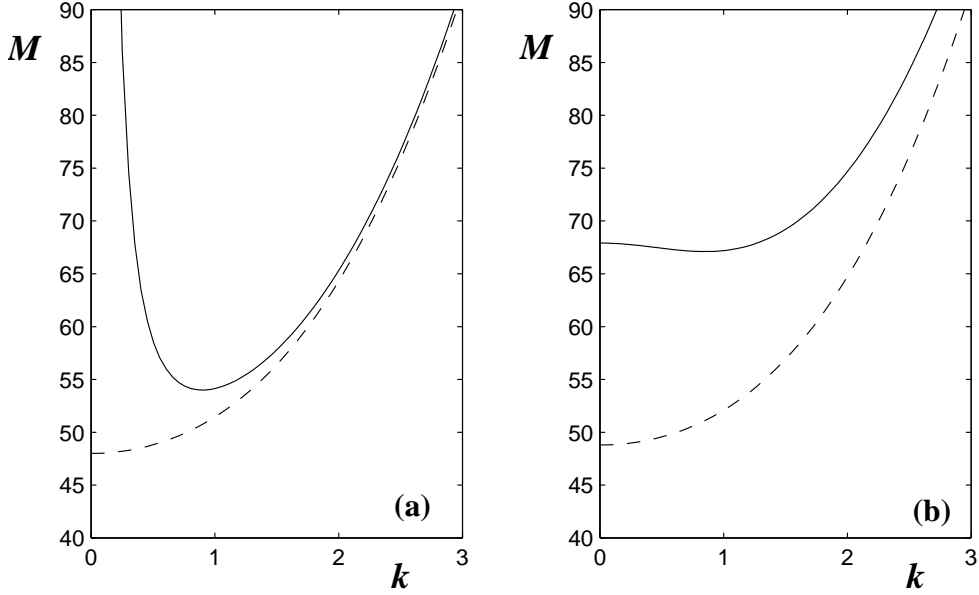,width=13cm}
\caption{(a) Linear stability curve obtained from 
the full governing equations without rotation for $Bi_1=Bi_2=0$
(dashed line) and for $Bi_1>0$ and $Bi_2>0$ (solid line); (b) Linear
stability analysis of the full governing equations  for $Bi_1=Bi_2=0$ 
at $\Omega=1$ (dashed line) and $\Omega=5$ (solid line).}
\LB{fig:linstab}
\end{figure}

As shown in Fig.\ref{fig:linstab}a the critical wavenumber $q_c$ vanishes for 
$Bi_1=Bi_2=0$, while for $Bi_1>0$ and $Bi_2>0$ one has $q_c>0$.   A
series expansion of (\ref{e:linstab}) around $Bi_1=Bi_2=0$ shows that
$q_c=((Bi_1+Bi_2) 405/27)^{1/4}$. Thus, for poorly conducting boundary
conditions the wavenumber of the growing pattern is still small enough to
allow the description by a long-wave model.  In Fig.\ref{fig:linstab}b the
marginal curve is shown for  $Bi_1=Bi_2=0$ and two values of the rotation
rate, $ \Omega = 1$ and $\Omega=5$. We see  that small values of the
rotation rate  do not modify the critical  wavenumber. However for 
$\Omega>\Omega_c\equiv3.7513$ a minimum develops in the neutral curve even 
for $Bi_1=Bi_2=0$ and the critical 
wavenumber changes sharply from zero to a finite value. Thus,  our study
will be valid  only for $\Omega<\Omega_c$. In this regime 
the critical Marangoni number at $q=0$ as a function of $\Omega$ is given by
\bea
M_c(\Omega)|_{q=0}={\frac {4\,{\Omega}^{3/2}\left(2\,\cos(\sqrt {\Omega})
\sin(\sqrt {\Omega})-\sinh(2
\,\sqrt {\Omega})\right )}{4\,\sinh(\sqrt {\Omega})\sin(\sqrt {\Omega})+\cos(2\,
\sqrt {\Omega})-\cosh(2\,\sqrt {\Omega})}}.
\LB{eq:Mc}
\eea

\section{Derivation of the Long-Wave Equation}
\LB{sec:longwave}

Here we derive the long-wavelength model for rotating  Marangoni
convection and consider the system  (\ref{e:feT})-(\ref{e:4}) together with the
boundary conditions (\ref{e:5})-(\ref{e:9}). Expanding $\theta$ and
${\bf u}$ around  the conductive solution, we rewrite the equations in terms
of the deviations from the basic solution. To avoid excessive notation we
keep the same notation $\theta$ and ${\bf u}$ for {\it deviations} of the
temperature and velocity fields and obtain
\bea 
\nabla^4 w - 2 \Omega \partial_z \zeta &=&0  \LB{eq:1},\\ 
\nabla^2 \zeta + 2 \Omega \partial_z w &=&0,  \LB{eq:2} \\
\Delta_H u +\partial_x \partial_z w + \partial_y \zeta &=&0,\LB{eq:3} \\
\Delta_H v + \partial_y \partial_z w - \partial_x \zeta  &=&0,\LB{eq:4}\\
\left( \partial_t  + {\bf u} \cdot \nabla \right) \theta &=&w+\Delta \theta.\LB{eq:6}
\eea
The boundary conditions are:
\bea
\theta_z -Bi_1 \theta=w=\zeta=w_z=0, &&  {\rm at} \, z=0, \LB{eq:7}\\
w_{zz}-M \Delta_H \theta= \theta_z + Bi_2 \theta =w=\zeta_z=0, & & {\rm at} \, z=1. \LB{eq:8}
\eea
We define a small parameter $\epsilon$
which measures distance from the critical Marangoni number by setting:
\bea
M=M_c(1+\epsilon^2 \mu ).\LB{eq:Mcexp}
\eea
Here $\mu$ is a parameter which allows to introduce changes
in the Marangoni number  independently of $\epsilon$.
We assume that the heat loss is small,
\bea
Bi=\epsilon^4 b.
\eea
As discussed above, this implies that the dimensionless wavelength
of the initial instability  is long, of order $\epsilon^{-1}$. This allow us
to introduce the slow scales,
\bea
X= \epsilon x, \,\,\,\,\,\,\, Y=\epsilon y.
\eea
The vertical coordinate $z$ is not rescaled. 
The evolution occurs on the slow time $T$, 
 \bea
 t=\epsilon^4 T,
 \eea
as suggested by the linear stability analysis \cite{Si82}. These scalings of
space and time suggest the scaling of the fields involved. Specifically, the
Marangoni condition (\ref{eq:8}) suggests that the vertical velocity be 
taken to be small while the deviation of the temperature from the linear profile
can be large,
\bea
(w,\theta)=(\epsilon^2 W,\Theta).
\eea
 Equations (\ref{eq:3}) and (\ref{eq:4}) point out the 
scaling for the $x$ and $y$ component of the velocity,
\bea
(u,v)=(\epsilon U, \epsilon V).
\eea
 From the definition of the $z$-component of the vorticity and considering
 the previous scaling we have,
 \bea
 \zeta= \epsilon^2 Z.
 \eea
With these definitions the equations (\ref{eq:1})-(\ref{eq:6}) become:
\bea
{\rm D}^4 W + 2 \epsilon^2 {\rm D}^2 \Delta_H W + \epsilon^4 \nabla_H^4 - 2
\Omega {\rm D} Z &=&0,\LB{equ:1}\\
{\rm D}^2 Z + 2 \Omega {\rm D} W + \epsilon^2 \Delta_H Z&=&0,\LB{equ:2}\\
\Delta_H U   + W_{Xz} + Z_Y&=&0,\LB{equ:3}\\
\Delta_H V   + W_{Yz} - Z_X&=&0,\LB{equ:4}\\
 \epsilon^4 \Theta_T - {\rm D}^4  \Theta - \epsilon^2 \Delta_H \Theta - 
\epsilon^2 W   &=&  
 - \epsilon^2 U \Theta_X- \epsilon^2 W \Theta_Y - \epsilon^2 W {\rm D} \Theta,\LB{equ:5}
 \eea
with D$\equiv \partial/\partial z$. The boundary conditions are,
\bea
W={\rm D} W = Z = {\rm D} \Theta - b_1 \epsilon^4 \Theta = 0, && \,\,{\rm on }\, z=0,\LB{equ:6}\\
W={\rm D}^2 W - M_c(1+\epsilon^2 \mu)\Delta_H \Theta = {\rm D} Z  
  = {\rm D} \Theta - b_2 \epsilon^4 \Theta = 0, && \,\,{\rm on }\, z=1.\LB{equ:7}
\eea
Proceeding  as in Refs. \cite{Kn90,Si82}, we expand the temperature as
follows,
 \[\Theta=\Theta_0(X,Y,z,T)+ \epsilon^2 \Theta_1 (X,Y,z,T)+...,\]
and  similarly for the other fields. Because in (\ref{equ:1})-(\ref{equ:7}) only even powers of $\epsilon$
appear, the expansion of the fields can also be restricted to even powers.
  If we insert these expansions in
(\ref{equ:1})-(\ref{equ:7}) we obtain  at ${\mathcal {O}}(\epsilon^0)$
\bea
\Theta_0&=&F(X,Y,T), \\
W_0&=&M \, \Delta_H F \,  f_1(\Omega,z), \\
Z_0&=&M \, \Delta_H F \, f_2(\Omega,z), \\
U_0&=&-M \, ( F_X \, {\rm D}f_2+F_Y \, f_1 ), \\
V_0&=&-M \, ( F_Y \, {\rm D}f_2-F_X \, f_1 ).  
\eea
In contrast to the non-rotating case the vertical vorticity $Z$ does
not vanish and at each order  we must solve two coupled
inhomogeneous differential equations instead of only one as in
\cite{Kn90}.  In these expressions $F(X,Y,T)$ is the planform function
whose governing equation we are looking for. The functions $f_1 $ and
$f_2$ are complicated functions of $\Omega$ and $z$. In the limit $\Omega
\to 0$ the function $f_2$ vanishes and  $f_1 $ is the same as in \cite{Kn90}.

At ${\mathcal{O}}(\epsilon^2)$ we obtain,
\[
{\rm D^2} \Theta_1 = - \Delta_H F - M \, f_1 \, \Delta_H F  -
 M \, {\rm D} f_1 \, | \nabla_H F|^2.
\]
The critical Marangoni number at $k=0$ is obtained from the solvability
condition  in agreement with the expression (\ref{eq:Mc}) obtained in the
linear stability analysis.  For the fields we obtain
 \bea
 \Theta_1&=& f_3(\Omega,z)  \Delta_H F + f_4 (\Omega,z)  | \nabla_H F|^2    + F_2(X,Y,T), \\
 W_1&=& f_5(\Omega,z)  \Delta_H F + f_6(\Omega,z) \Delta_H^2 F + f_7(\Omega,z) 
 \Delta_H | \nabla_H F|^2 + \nonumber \\
 &&  f_8(\Omega,z) \Delta_H F_2, \\
 Z_1 &=& f_9(\Omega,z)  \Delta_H F + f_{10}(\Omega,z) \Delta_H^2 F 
 + f_{11}(\Omega,z) 
 \Delta_H | \nabla_H F|^2 +\nonumber \\
 &&  f_{12}(\Omega,z) \Delta_H F_2, \\
 U_1 &=& -f_9 \,  F_Y - f_{10} \,\Delta_H  F_Y - f_{11} \,
 (| \nabla_H F|^2)_{Y} - f_{12} \, F_{2, Y} - f_{5, z} \,  
  F_X - \nonumber \\ && f_{6, z} \, \Delta_H  F_X -  
   f_{7, z} 
 (| \nabla_H F|^2)_{X} - f_{8, z}   F_{2, X},\\
 V_1 &=& f_9 \,  F_X + f_{10} \,\Delta_H  F_X + f_{11} \,
 (| \nabla_H F|^2)_{X} + f_{12} \, F_{2, X} - f_{5, z} \,   
 F_Y - \nonumber \\ && f_{6, z} \, \Delta_H  F_Y -  
  f_{7, z} 
 (| \nabla_H F|^2)_{Y} - f_{8, z}   F_{2, Y}.
 \eea
Here $F_2$ is the $O(\epsilon^2)$-correction to the planform
function $F$.  The functions $f_3$, $f_4$, $f_5$, $f_6$, $f_7$, $f_8$,  
$f_9$, $f_{10}$,  $f_{11}$ and $f_{12}$  are again complicated
functions   of $z$ and $\Omega$. The functions $f_3$ and $f_4$ are
related   to integrals of $f_1$ with respect to $z$. It turns out that after
the   integration  these functions contain terms that are singular at  
$\Omega=0$ and  do not depend on $z$ in this limit. In order to  
cancel these singularities and to recover the expressions of  
\cite{Kn90} in the limit $\Omega \to 0$, the planform function $F_2$,
which is undetermined at this order, must contain the same singular
constants with the opposite sign. Since the integral expressions are
too complicate to  present here, this is best illustrated by the simple
example
  \bea
  \int \sin(\Omega z) dz = - \frac{1}{\Omega}\cos(\Omega z) + C.
  \eea
  This integral is not finite in the limit $\Omega \to 0$. If 
  a finite value is required for it, the arbitrary constant $C$ must be 
  chosen as $1/\Omega$. Analogously, in $f_3$ and $f_4$ constants
appear that need to be chosen appropriately. Since they do not depend on $z$ they 
can be absorbed in $F_2$ and therefore do not affect the long-wave equation for $F$. 

 \begin{figure}[tbp]  
  \epsfig{file=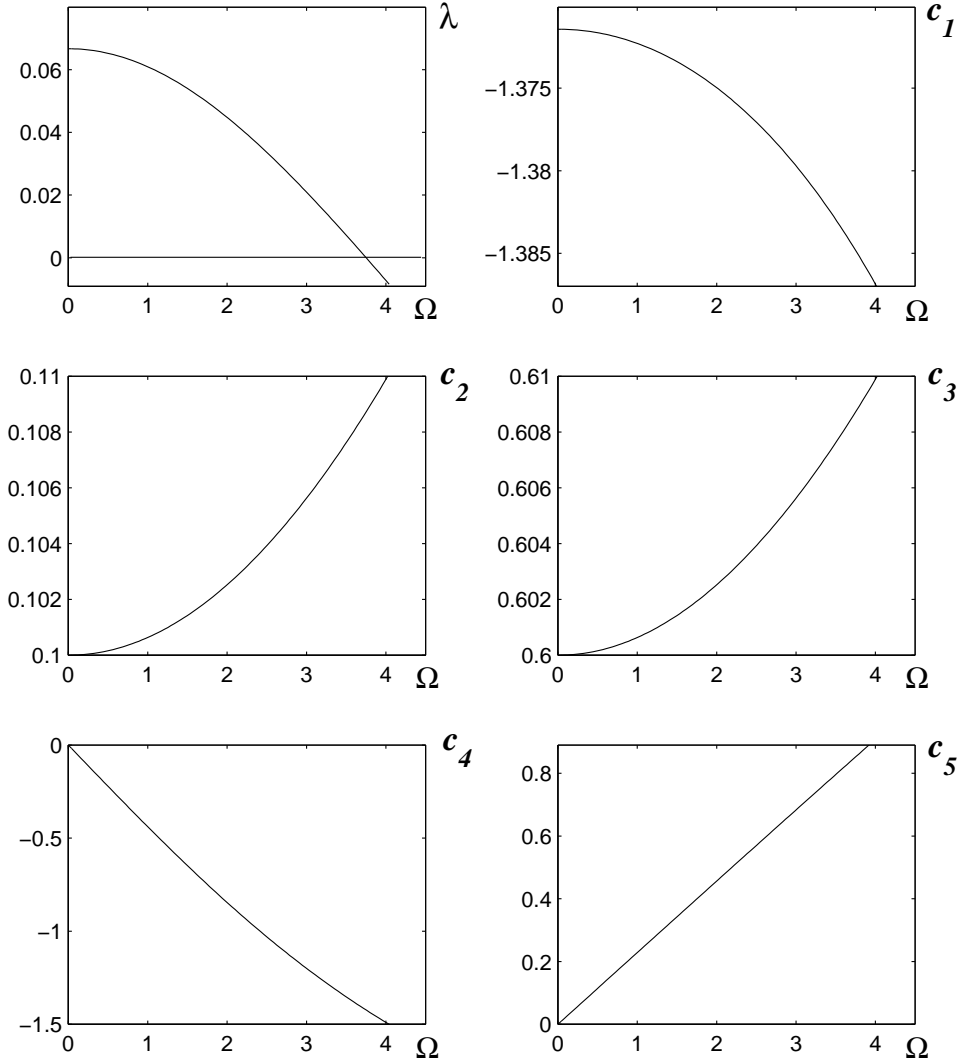,width=13cm}  
\caption{The constants $\lambda$, $c_1$, $c_2$, $c_3$, $c_4$ and $c_5$   
  plotted as functions of the rotation rate $\Omega$.}  
\end{figure}  
 
  Finally, at ${\mathcal{O}} (\epsilon^4)$ the 
  solvability condition for D$^2 \Theta_2$ yields the required equation for  
  the planform function $F(X,Y,T)$,
  \bea 
  F_T&=& - (b_1+b_2) F - \mu \Delta_H F - \lambda \Delta_H^2 F - 
  c_1 \nabla_H \cdot | \nabla_H F|^2
  \nabla_H F - \nonumber \\
  & & c_2 \nabla_H \cdot \Delta_H F \nabla_H F - c_3 \Delta_H | \nabla_H F|^2 
-c_4 {\bf e}_z \cdot \nabla_H F \times \nabla_H \left\{\Delta F\right\} \nonumber \\
& &+c_5 {\bf e}_z \cdot\nabla_H F \times \nabla_H \left\{(\nabla_H F)^2\right\}.
\LB{eq:lwm}
\eea  
  Here $\mu$ is the control parameter (cf. (\ref{eq:Mcexp})) 
   The constants $\lambda$, $c_1$, $c_2$, $c_3$, $c_4$ and $c_5$ 
  are complicated functions of the rotation rate $\Omega$. We plot
  them in  Fig. 2. A series expansion of these functions around $\Omega=0$,
  shows that in the limit $\Omega \to 0$, the same equation  as
  in \cite{Kn90} is obtained:
  \bea
  \lambda&=&{\frac {1}{15}}-{\frac {7}{1200}}{\Omega}^{2}+\mathcal{O}\left
({\Omega}^{4}
 \right ),           \nonumber \\
  c_1&=&-{\frac {48}{35}}-{\frac {104}{121275}}{\Omega}^{2}+\mathcal{O}\left ({\Omega}^
{4}\right ),               \nonumber \\
  c_2&=&{\frac {1}{10}}+{\frac {1}{1575}}{\Omega}^{2}+\mathcal{O}\left ({\Omega}^{4}
\right ),               \nonumber \\
  c_3&=&{\frac {3}{5}}+{\frac {1}{1575}}{\Omega}^{2}+\mathcal{O}\left ({\Omega}^{4}
\right ),
   \nonumber \\
  c_4&=&-{\frac {233}{525}}\Omega+{\frac {404}{72765}}{\Omega}^{3}+\mathcal{O}\left ({
\Omega}^{5}\right ),   \nonumber \\
  c_5&=&{\frac {8}{35}}\Omega-{\frac {122}{1576575}}{\Omega}^{3}+\mathcal{O}\left ({
\Omega}^{5}\right ).
       \nonumber  
  \eea

\section{Numerical results}
\LB{sec:num}

We now turn to the stability and dynamics of hexagon patterns described by
(\ref{eq:lwm}). One of our goals is to identify parameter regimes in which
regular hexagon patterns  become unstable at all  wavenumbers. In the
weakly nonlinear case hexagons can be described by  three coupled
Ginzburg-Landau equations. Within this framework their stability and
dynamics has been studied in detail without and with rotation
\cite{LaMe93,SuTs94,EcPe98,NuNe00,EcRi00,EcRi00a,YoRi02}.
It has been found that for small amplitudes there is always a range of
wavenumbers over which the hexagons are stable.   In a simple
Swift-Hohenberg model, situations were  identified in which somewhat further
above onset the stability balloon closes, leading to persistent chaotic
dynamics \cite{SaRi00}. In this paper we therefore do not reduce
(\ref{eq:lwm}) to small-amplitude Ginzburg-Landau equations. Instead
determine the stability of the hexagon solutions of (\ref{eq:lwm}) and the
resulting dynamics numerically without any restriction to small amplitudes.

The stability of the hexagon patterns is determined with a Fourier-Galerkin
code on a hexagonal lattice. We then determine their stability with respect to
side-band perturbations using a  Floquet analysis with a Floquet parameter
${\bf p}=p(\cos \theta, \sin \theta)$. Due to the isotropy of the system and the discrete 
rotational symmetry of the hexagons we can
restrict our analysis to perturbation wavevectors subtending an angle
$\theta$ of up to $\pm 30^o$  relative to the hexagon modes. It turns out that 
accurate solutions of (\ref{eq:lwm}) require a large number of Fourier modes.
One measure of the accuracy of a solution is obtained with the stability
analysis by determining the magnitude of the eigenvalue corresponding to
the translation mode given by the  Floquet parameter ${\bf p}=0$. For the
exact solution one has $\sigma ({\bf p})=0$, while numerically truncated
solutions will have  $\sigma ({\bf p})\ne 0$. The dependence of $\sigma ({\bf
p})$ as a function of the number $n$ of harmonics retained is illustrated in
table \ref{t:sigman}, where $n=1$  corresponds to the basic hexagon solution.

\begin{table}
\begin{center}
\begin{tabular}{|l|l|l|l|l|l|}\hline
$\mu$ & $\Omega$ & q & $n=7$ & $n=11$ & $n=15$\\ \hline
$0.07$ & $3.0$ & $0.98$ & $-0.16 \cdot 10^{-3}$ & $-0.28\cdot 10^{-5}$ & $-0.56\cdot 10^{-7}$ \\ \hline
$0.17$ & $2.5$ & $0.95$ & $-0.25 \cdot 10^{-1}$ & $-0.53\cdot 10^{-2}$ & $-0.60\cdot 10^{-3}$ \\ \hline
\end{tabular} 
\caption{Translation eigenvalue $\sigma ({\bf p}=0)$ as a function of the truncation parameter $n$
of the Fourier expansion for $b_1+b_2=0.025$ illustrating the poor convergence
for larger amplitudes.\LB{t:sigman}}
\end{center}

\end{table}

Fig.\ref{fig:stabom0} shows the resulting stability limits in the absence of
rotation ($\Omega=0$) for $b_1+b_2=0.025$. They are given by steady long-wave instabilities 
with $p \ll 1$  (cf.\cite{LaMe93,SuTs94,YoRi02}). Numerical simulations for
selected parameter values (marked by plusses and crosses in
fig.\ref{fig:stabom0}) confirm that the instabilities lead to the formation of
defects which then change the wavevector of the pattern. 
 
\begin{figure}[tbp]  
  \epsfig{file=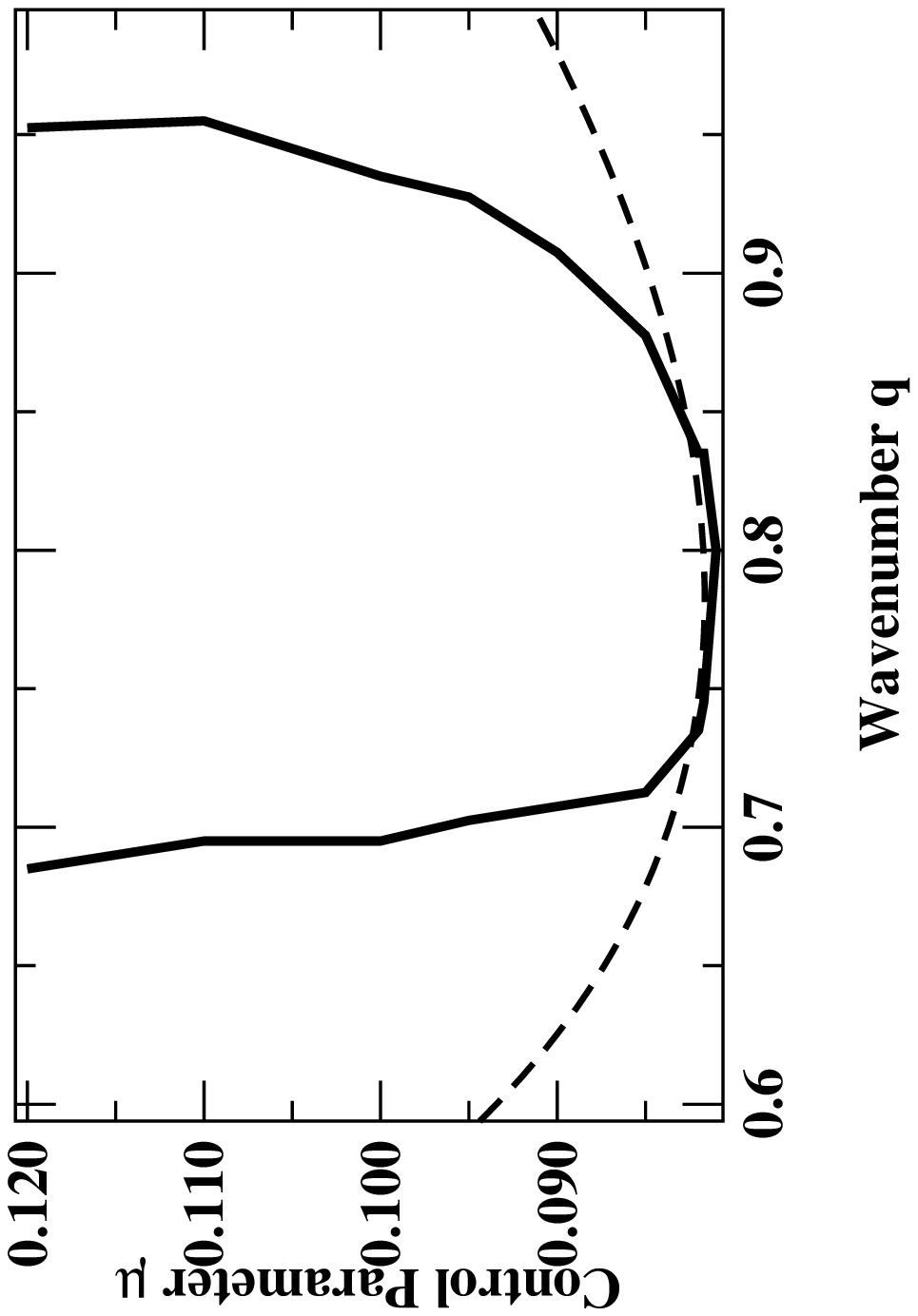,width=13cm}  
\caption{Stability limits of the steady hexagons for $\Omega=0$ and $b_1+b_2=0.025$. The dashed
line gives the neutral curve. The stability limit is given by a steady long-wave
instability (solid line). The plusses and crosses indicate stable and unstable
solutions according to the numerical simulation.
}
\LB{fig:stabom0} 
\end{figure}

Rotation strongly modifies the stability limits as shown  in fig.\ref{fig:stabom3}
for $\Omega=3$ and $b_1+b_2=0.025$. The most striking aspect is that the stability balloon closes
for $\mu>0.07$. This closing is due to a steady short-wave instability
(triangles) which supersedes the usual steady long-wave instability (open circles)
for larger $\mu$. In the short-wave regime the largest growth rates are
obtained for $p\approx 0.6 q$ and $\theta \approx 25^o-30^o$.  The
appearance of a short-wave instability  is somewhat similar to the result
found in the weakly nonlinear analysis. There it has been found that in the
presence of rotation a steady or oscillatory instability arises at finite
modulation wavenumbers and preempts the long-wave oscillatory instability
that emerges from the interaction of the two long-wave phase modes
\cite{EcRi00}. 

\begin{figure}[tbp]  
  \epsfig{file=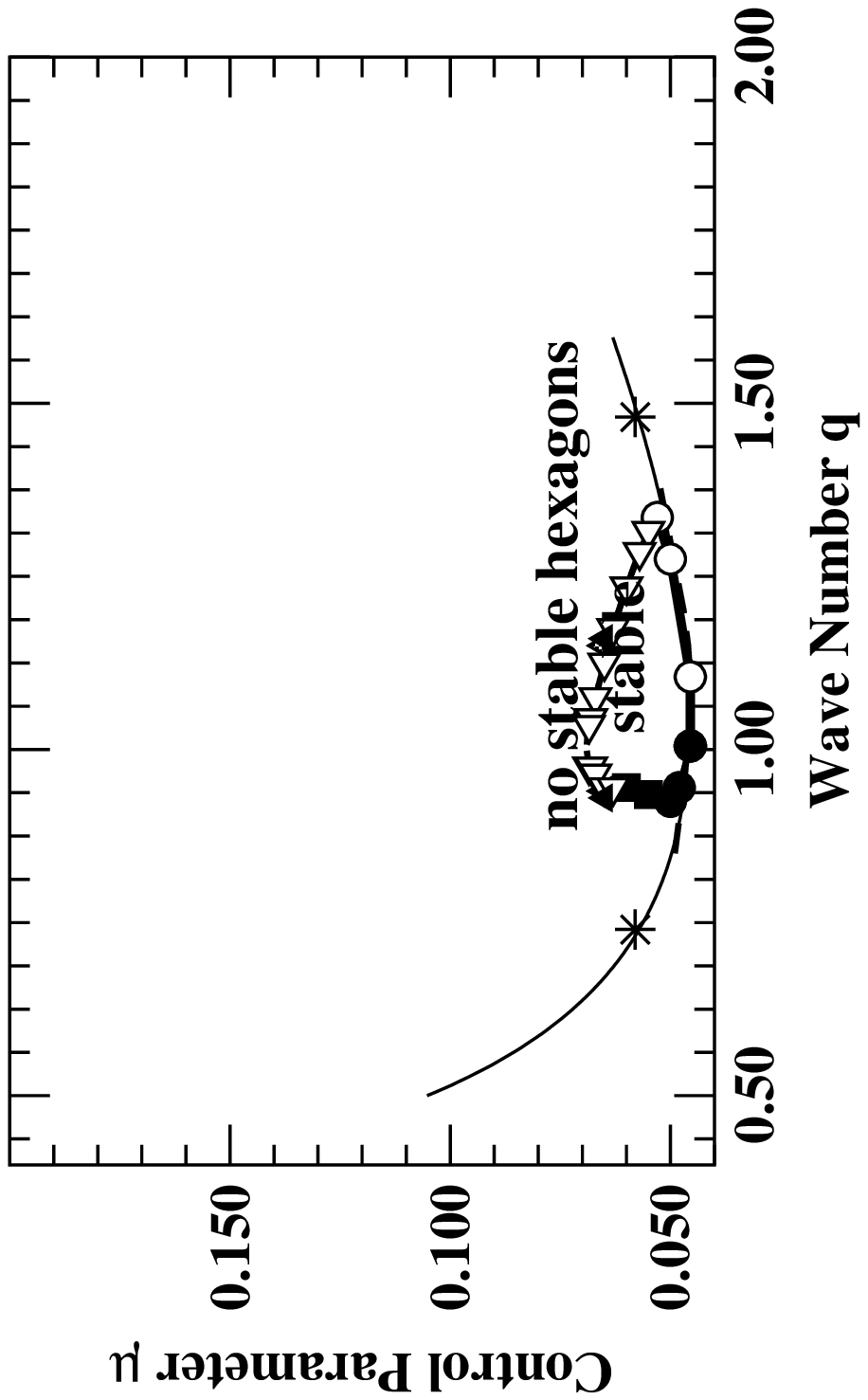,width=13cm}  
\caption{Stability limits of the steady hexagons for $\Omega=3$ and $b_1+b_2=0.025$. Hexagons
are only stable inside the closed loop. The neutral and saddle-node lines
are given by thin dashed and solid lines. The symbols identify the
 character of the relevant instabilities: long-wave steady  (open circles),
short-wave steady (open left-triangles), long-wave oscillatory (solid circles),
homogeneously oscillatory (solid squares). The plusses and crosses
denote simulations in which the steady hexagons were stable/unstable (see text).
} 
\LB{fig:stabom3} 
\end{figure}

A further interesting feature of the stability limits for $\Omega=3$ is the
oscillatory instability that limits the stability balloon on the low-$q$ side (solid
symbols). For low values of $\mu$ it is a long-wave side-band instability,
i.e. its growth rate  and frequency vanish for ${\bf p}=0$ (solid circles). For
larger $\mu$, however, the largest growth rate occurs at ${\bf p}=0$ and the
hexagon pattern is unstable to a homogeneous oscillatory mode (solid
squares). In the weakly nonlinear analysis such a homogeneous oscillatory 
instability has only been identified in the context of the transition from
hexagons  to rolls \cite{Sw84,So85,EcRi00}. No such connection is apparent
in the present system.

Anticipating interesting dynamics for values of $\mu$ above the closing of
the stability balloon, we have studied its dependence on the rotation rate. As
a function of $\Omega$, fig.\ref{fig:mumax} shows the maximal control
parameter $\mu$ for which there are still stable steady hexagons (solid squares).
To indicate the $\mu$-range of stable hexagons the critical control parameter $\mu_c$ 
is also given (dashed line). With decreasing $\Omega$ that $\mu$-range grows rapidly.
To resolve the stability limits for yet smaller values of $\Omega$ 
a very large number of modes is needed;  $n=19$ is not sufficient (cf. table \ref{t:sigman}).


\begin{figure}[tbp]  
  \epsfig{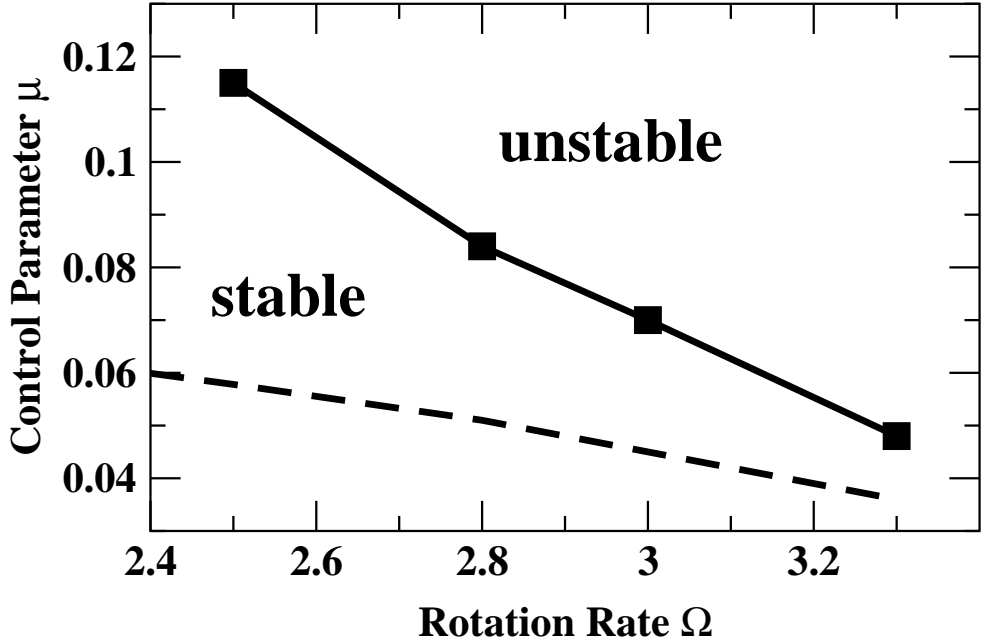}  
\caption{Maximal value of the control parameter for which there are still 
steady hexagons (solid line) and onset of hexagons (dashed line) ($b_1+b_2=0.025$).
} 
\LB{fig:mumax} 
\end{figure}

The nonlinear evolution ensuing from the linear instabilities is investigated by
simulating  (\ref{eq:lwm}) numerically with a pseudo-spectral Fourier code
using an integrating-factor, fourth-order Runge-Kutta scheme. It turns out
that for larger values of $\mu$ the nonlinear gradient terms in (\ref{eq:lwm})
can under certain conditions lead to an apparently spurious blow-up of the
numerical solution in which very
high wavenumbers are excited \cite{Sa00}\footnote{Even though the
numerical time-stepping routine may lead to a blow-up, no change in the
behavior of the stationary states is found with the Galerkin code for those
parameters.  Apparently, in these cases the Fourier representation of the
derivatives in the time-stepping code does not provide sufficient numerical
dissipation. By introducing an artificial dissipation term  these divergencies
can be avoided. To obtain sufficiently smooth starting solutions  we have at
times  used a filter that affects only wavenumbers above a certain high
wavenumber and corresponds then to a linear 8th-order derivative term.
However, all the simulations presented in this paper have been performed
with that filter turned off.}. However, if a sufficiently smooth initial condition
was used no such numerical divergencies occurred.  As already indicated
earlier (cf. table \ref{t:sigman}) the  proper resolution of the patterns requires a
large number of Fourier modes.  In all simulations we have used a
rectangular computational domain of size $L_X=2L$ and $L_Y=2L/\sqrt{3}$
to accommodate undeformed hexagon patterns. Specifically, for
$L=N\,2\pi/q$ this domain allows $N$ periods of the pattern in the $x$- and
$y$-direction.

To test the stability results we have performed simulations near the  stability
limits depicted in figs.\ref{fig:stabom0},\ref{fig:stabom3}. The plusses
denote parameters for which a small modulation of the  pattern did not
destabilize it, while at the crosses the hexagons  became unstable.
These simulations were performed in systems of size $L=4\cdot2\pi/q$. For
this aspect ratio some of the instabilities are still slightly suppressed shifting
the stability limits somewhat. The limits obtained in the simulations are,
however, consistent with the stability results if the modulation wavenumbers
${\bf p}$ are restricted to those compatible with the computational domain
used in the simulations. 

We first address the nonlinear evolution of the homogeneous oscillatory
instability (solid squares). Simulations for $\mu=0.060$ and $q=0.93$ show
that this stability limit is due to a supercritical Hopf bifurcation, i.e. the
spatially homogeneous oscillations saturate and lead to a state of oscillating
hexagons. They are similar in appearance to the oscillating hexagons  that
arise in the Hopf bifurcation that replaces the instability of hexagons to  the
mixed mode, which is associated with the transition to rolls. As mentioned
earlier, this transition to the mixed mode and to rolls has not been seen in the
present system. A time sequence of snapshots of the oscillating hexagons is
shown in fig.\ref{fig:oschex}\footnote{A movie of the temporal evolution of
the oscillating hexagons can be found at {\tt http://www.esam.northwestern.edu/riecke/research/Marangoni/marangoni.html}.}.

\begin{figure}[tbp]  
\begin{center}
see figures 
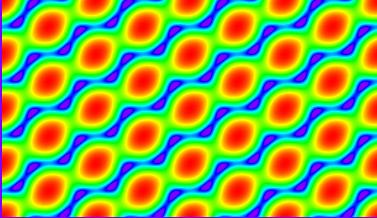
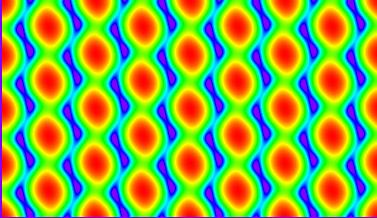
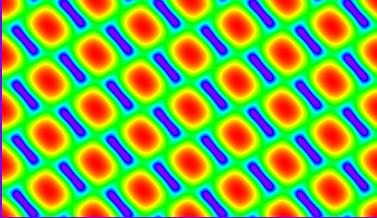
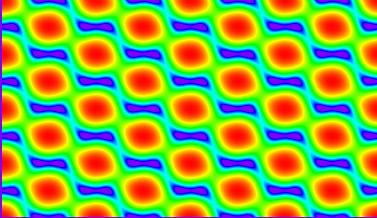
\end{center}
\caption{Snapshots of oscillating hexagons for $\Omega=3$ and $b_1+b_2=0.025$, 
$\mu=0.060$, and $q=0.93$ at times  
$t=0$, $t=\frac{1}{3}T$, $t=0.43T$, and $t=\frac{2}{3}T$, where
the period $T=140$ (time progressing left to right, top to  bottom; red denotes
positive, blue negative values). }
\LB{fig:oschex}  
\end{figure} 

Now we turn to the nonlinear evolution arising from the steady short-wave
instability (open left-triangles in fig.\ref{fig:stabom3}). In fig.\ref{fig:patt20} and
fig.\ref{fig:time20}   a sequence of the pattern and the temporal evolution of 
two modes is shown, respectively  in a system of size $L_X=20.8$ for
$\mu=0.07$ (using $128\times128$ modes). For this value of the  control
parameter no stable hexagon patterns exist in the infinite system. In this case
the temporal evolution  does not become periodic for the time period
investigated. 

   
\begin{figure}[tbp]  
\begin{center}
see 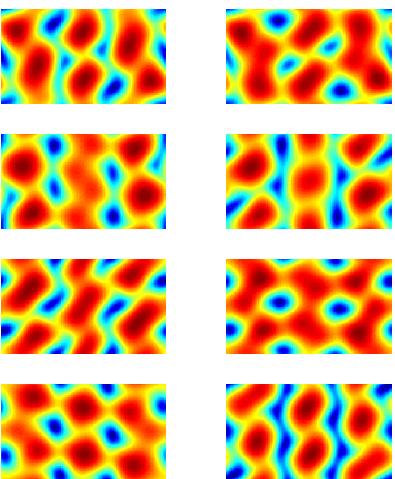 
\end{center}
\caption{ Snapshots of typical patterns obtained for $\Omega=3$ and $b_1+b_2=0.025$ at $L_X=20.8, \mu=0.07$ at 
times 6299, 6349, 6449, 6499, 6549, 6599, 6649, 6749 (left to right and top to bottom)}  
\LB{fig:patt20}
\end{figure}   
\begin{figure}[tbp]  
  \epsfig{file=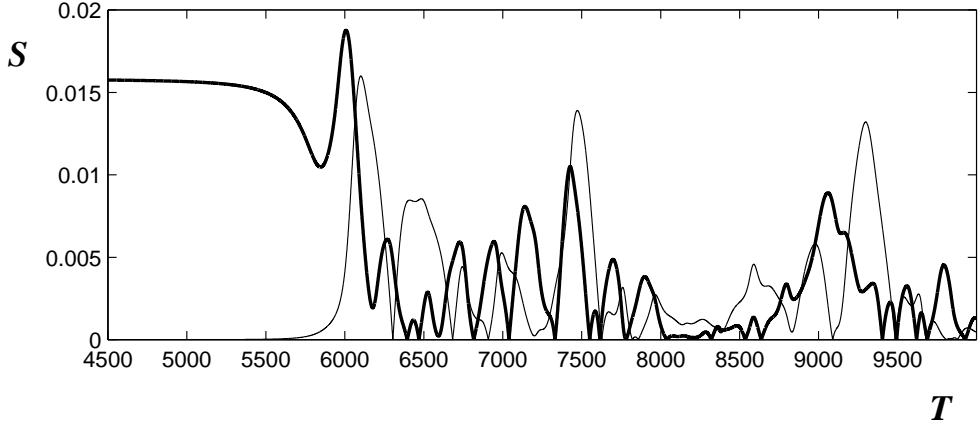,width=13cm}  
\caption{ Time series obtained for $\Omega=3$ and $b_1+b_2=0.025$ at 
$L_X=20.8$, $\mu=0.07$. It shows
 the energy $S$ associated with the Fourier modes with wavevectors 
$(2\pi/20.8,2\cdot2\pi\sqrt{3}/20.8)$ with thin and thick solid lines,
respectively. Note the long onset time for the instability.} 
\LB{fig:time20} 
\end{figure}
   
In fig.\ref{fig:patt48} typical snapshots for the evolution of the pattern in a
system of size $L_X=48$ are shown for $\mu=0.08$ (using $256\times256$
modes). The associated time series for two Fourier modes are shown in
fig.\ref{fig:time48}. Again, for these parameter values no steady stable
hexagons exist in large systems resulting in persistent chaotic
dynamics.\footnote{A movie of the chaotic dynamics for $\mu=0.075$ and
$L_X=48$ can be found at {\tt http://www.esam.northwestern.edu/riecke/research/Marangoni/marangoni.html}.} While the
pattern is spatially quite disordered a tendency to form locally square
arrangements can be observed. This presumably reflects the fact that in the
absence of the quadratic terms in (\ref{eq:lwm}) even in the rotating system
square patterns are preferred over roll patterns \cite{Co98}.
 
\begin{figure}[tbp]  
\begin{center}
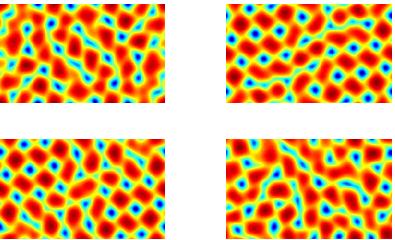
\end{center}
\caption{Typical patterns obtained for $\Omega=3$ and $b_1+b_2=0.025$ at $L_X=48, \mu=0.08$ at times
 2699, 3749, 3899, 11549 (left to right and top to bottom).}  
\LB{fig:patt48}
\end{figure}   
 
\begin{figure}[tbp]  
\begin{center}
see 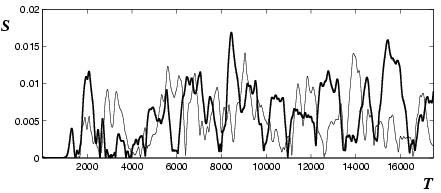
\end{center}
\caption{ Time series obtained for $\Omega=3$ and $b_1+b_2=0.025$ at $L_X=48, \mu=0.08$. The energy $S$
associated with the amplitude of the Fourier modes with wavevector
$(2\cdot 2\pi/48,4\cdot 2\pi\sqrt{3}/48)$ and $(0,4\cdot 2\pi\sqrt{3}/48)$
are shown with thin and thick solid lines, respectively.}
\LB{fig:time48}  
\end{figure} 
 
\begin{figure}[tbp]  
\begin{center}
see 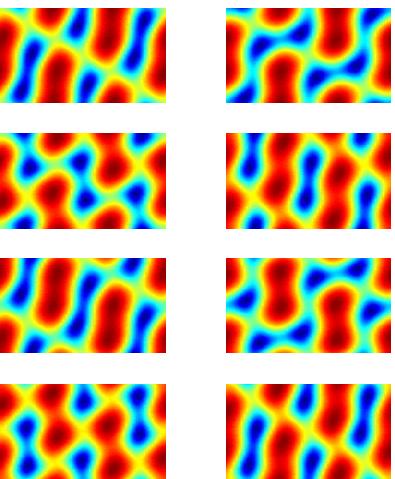
\end{center}
\caption{ Patterns in a cycle obtained for $\Omega=3$ and $b_1+b_2=0.025$ at $L_X=17$, $\mu=0.058$ at times
4079.2, 4159.2, 4239.2, 4319.2, 4399.2, 4479.2, 4559.2, 
4639.2 (left to right and top to bottom).}  
\LB{fig:patt17}
\end{figure}  
 

\begin{figure}[tbp]  
\epsfig{file=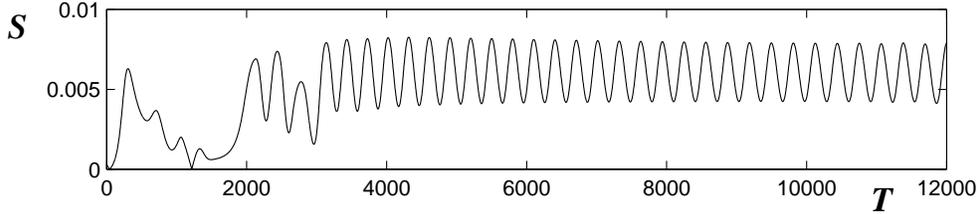,width=13cm}  
\caption{ Time series obtained for $\Omega=3$ and $b_1+b_2=0.025$ at $L_X=17$, $\mu=0.058$. It shows the 
energy $S$ associated with the Fourier mode with wavevector 
$(2\pi/17,2\cdot 2\pi\sqrt{3}/17)$.
} 
\LB{fig:time17}  
\end{figure} 

Finally, in Figs.\ref{fig:patt17}, \ref{fig:time17} we show the temporal
evolution in a quite small system of  size $L_X=17$ for $\mu=0.058$
(using $64\times64$ Fourier modes). For such a small system the
stability limits as shown in figs.\ref{fig:stabom3} do not really apply.
The initial condition was a regular hexagon  pattern with wavenumber
$q=1.478$.  After a transient during which defects appeared and
changed the number of cells the time dependence of the pattern
becomes quite regular. The time series  (cf. fig.\ref{fig:time17})  is,
however, not strictly periodic but shows small ($\approx 1\%$)
irregular variations of the oscillation amplitude.  A closer inspection of
the neutral curve, which in terms of the system size and of the
number $N$ of wavelengths contained in the system is given by
\bea
\mu(L_x)=b\left(\frac{L_X}{4\pi N}\right)^2+\lambda \left(\frac{4\pi N}{L_X}\right)^2,
\LB{e:neutralL}
\eea
shows that the parameters for this simulation are close to the
codimension-two point at which hexagons with $q=0.74$ and hexagons with
$q=1.48$ first arise. This is illustrated in fig.\ref{fig:neutralL} which  gives the
neutral curve (\ref{e:neutralL}) for various values of $N$. In this small system
we associate the origin of the dynamics with the low-dimensional interaction
of the two modes associated with the codimension-two point. 

\begin{figure}[tbp]  
\epsfig{file=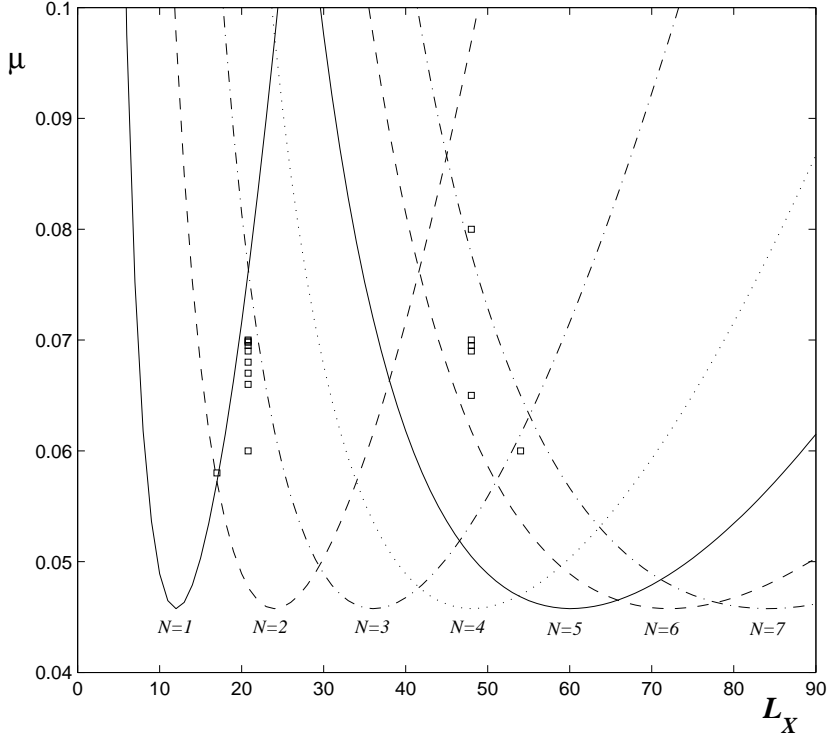,width=11cm}  
\caption{The neutral curve $\mu(L_X)$ plotted for different values of $N$, the 
number of wavelengths contained in the system. The  
parameters are $b_1+b_2=0.025$ and $\lambda=0.0209$ (which corresponds  
to a rotation rate of $\Omega=3$). 
Squares stand  for numerical simulations performed in different boxes at different $\mu$ values.}
\LB{fig:neutralL}  
\end{figure} 

\section{Conclusion}

In this paper we have considered Marangoni convection in the presence of
rotation. In order to derive systematically a  set of reduced equations that -
unlike the Ginzburg-Landau equations - preserve the  isotropy of the system
we have considered poorly conducting boundary conditions. Under these
conditions convection sets in first at long wavelengths, which allows the
derivation of a long-wave equation. Extending previous
analyses \cite{Kn90,ShSi91}, we have derived the long-wave equation 
in the presence of rotation and obtained the coefficients as a function 
of the rotation rate $\Omega$. Using a Galerkin
approach we have then determined the linear stability of steady hexagonal
convection with respect to  general side-band perturbations without and with
rotation. We have employed direct numerical  simulations of the long-wave
equation to investigate the fate of these instabilities. 

In the absence of rotation the band of stable steady hexagons in
large systems is determined solely by a steady long-wave instability
for the range of the control parameter considered. With rotation,
however,  steady long- and short-wave instabilities as well as
oscillatory instabilities  arise. The latter can be a long-wave
side-band instability or an instability that is spatially homogeneous at
its onset. Interestingly, the latter is supercritical and  leads to stable,
spatially homogeneous oscillations of the hexagons. This state
resembles the oscillating-hexagon branch that arises in
non-Boussinesq buoyancy-driven convection in connection with the
transition from hexagon to roll convection \cite{Sw84,So85,EcRi00}. 
In the present system no transition to rolls has been identified and in
experiments on Marangoni convection a  transition from hexagons to
squares is observed instead \cite{NiTh95,ScVa99}. We find
temporally almost periodic and spatially more complex states that
appear to arise from a resonant $q:2q$ mode interaction. This and
other resonances could be studied in future work.   Most interesting is
the fact that for sufficiently large rotation rates the stability limits close
for larger values of the control parameter. In the numerical simulations
this is shown to lead to spatially and temporally  chaotic patterns. 

 So far,  there have been very few experimental studies of spatio-temporal
chaos arising from a  hexagonal planform \cite{OuSw91}. Therefore, it would
be particularly interesting to see whether the complex dynamics of the 
disordered hexagonal patterns predicted by the long-wave model
(\ref{eq:lwm}) can be obtained in experiments. It is yet to be seen to what
extent in the  current system the deformation of the free surface by the
centrifugal force modifies the results presented here. Similarly, the derivation
of the long-wave equation  depends on the poor thermal conductivity of the
horizontal boundaries. For  experimentally realizable boundary conditions
the wavelength of the convecting state may not be as large \cite{VaSc97}. 
As is often the case, however, the instabilities predicted by the long-wave
equation (\ref{eq:lwm}) may persist qualitatively beyond the asymptotic
long-wave limit. A study of this question will presumably require the 
derivation of the Ginzburg-Landau equations with nonlinear gradient terms
or, more likely due to the finite amplitudes at the closing of the stability 
balloon, a numerical analysis of the full fluid equations (cf.\cite{Be93}).

\begin{ack}
AMM thanks the Gobierno de Navarra for a Postdoctoral Fellowship
and the Universidad de Navarra for partial support. Part of this work 
was done during a  visit by AMM at Northwestern University, which
she would like to thank for its hospitality and support. We would also like
to thank the Centro de Astrobiolog{\'{\i}}a and INTA for
providing access to  their computer systems where most of the
numerical simulations were performed. HR gratefully acknowledges
support by NASA (NAG3-2113), the Engineering Research Program
of the Office of Basic Energy Sciences at the Department of Energy
(DE-FG02-92ER14303), and by a grant from NSF (DMS-9804673).

\end{ack}

\bibliography{journal}

\end{document}